
\input harvmac.tex


\def\unlockat{\catcode`\@=11}

\def\lockat{\catcode`\@=12}

\unlockat


\def\newsec#1{\global\advance\secno by1\message{(\the\secno. #1)}
\global\subsecno=0\global\subsubsecno=0
\global\deno=0\global\prono=0\global\teno=0\eqnres@t\noindent
{\bf\the\secno. #1}
\writetoca{{\secsym} {#1}}\par\nobreak\medskip\nobreak}
\global\newcount\subsecno \global\subsecno=0
\def\subsec#1{\global\advance\subsecno
by1\message{(\secsym\the\subsecno. #1)}
\ifnum\lastpenalty>9000\else\bigbreak\fi\global\subsubsecno=0
\global\deno=0\global\prono=0\global\teno=0
\noindent{\it\secsym\the\subsecno. #1}
\writetoca{\string\quad {\secsym\the\subsecno.} {#1}}
\par\nobreak\medskip\nobreak}
\global\newcount\subsubsecno \global\subsubsecno=0
\def\subsubsec#1{\global\advance\subsubsecno by1
\message{(\secsym\the\subsecno.\the\subsubsecno. #1)}
\ifnum\lastpenalty>9000\else\bigbreak\fi
\noindent\quad{\secsym\the\subsecno.\the\subsubsecno.}{#1}
\writetoca{\string\qquad{\secsym\the\subsecno.\the\subsubsecno.}{#1}}
\par\nobreak\medskip\nobreak}

\global\newcount\deno \global\deno=0
\def\de#1{\global\advance\deno by1
\message{(\bf Definition\quad\secsym\the\subsecno.\the\deno #1)}
\ifnum\lastpenalty>9000\else\bigbreak\fi
\noindent{\bf Definition\quad\secsym\the\subsecno.\the\deno}{#1}
\writetoca{\string\qquad{\secsym\the\subsecno.\the\deno}{#1}}}

\global\newcount\prono \global\prono=0
\def\pro#1{\global\advance\prono by1
\message{(\bf Proposition\quad\secsym\the\subsecno.\the\prono #1)}
\ifnum\lastpenalty>9000\else\bigbreak\fi
\noindent{\bf Proposition\quad\secsym\the\subsecno.\the\prono}{#1}
\writetoca{\string\qquad{\secsym\the\subsecno.\the\prono}{#1}}}

\global\newcount\teno \global\prono=0
\def\te#1{\global\advance\teno by1
\message{(\bf Theorem\quad\secsym\the\subsecno.\the\teno #1)}
\ifnum\lastpenalty>9000\else\bigbreak\fi
\noindent{\bf Theorem\quad\secsym\the\subsecno.\the\teno}{#1}
\writetoca{\string\qquad{\secsym\the\subsecno.\the\teno}{#1}}}
\def\subsubseclab#1{\DefWarn#1\xdef
#1{\noexpand\hyperref{}{subsubsection}%
{\secsym\the\subsecno.\the\subsubsecno}%
{\secsym\the\subsecno.\the\subsubsecno}}%
\writedef{#1\leftbracket#1}\wrlabeL{#1=#1}}

\lockat


\def\IB{\relax\hbox{$\inbar\kern-.3em{\rm B}$}}
\def\IC{\relax\hbox{$\inbar\kern-.3em{\rm C}$}}
\def\ID{\relax\hbox{$\inbar\kern-.3em{\rm D}$}}
\def\IE{\relax\hbox{$\inbar\kern-.3em{\rm E}$}}
\def\IF{\relax\hbox{$\inbar\kern-.3em{\rm F}$}}
\def\IG{\relax\hbox{$\inbar\kern-.3em{\rm G}$}}
\def\IGa{\relax\hbox{${\rm I}\kern-.18em\Gamma$}}
\def\IH{\relax{\rm I\kern-.18em H}}
\def\IK{\relax{\rm I\kern-.18em K}}
\def\IL{\relax{\rm I\kern-.18em L}}
\def\IP{\relax{\rm I\kern-.18em P}}
\def\IR{\relax{\rm I\kern-.18em R}}
\def\IZ{\relax\ifmmode\mathchoice
{\hbox{\cmss Z\kern-.4em Z}}{\hbox{\cmss Z\kern-.4em Z}}
{\lower.9pt\hbox{\cmsss Z\kern-.4em Z}}
{\lower1.2pt\hbox{\cmsss Z\kern-.4em Z}}\else{\cmss Z\kern-.4em Z}\fi}

\def\II{\relax{\rm I\kern-.18em I}}

\def\frac#1#2{{#1\over#2}}




\def\ch{{\rm ch}}

\def\inbar{\,\vrule height1.5ex width.4pt depth0pt}
\font\cmss=cmss10 \font\cmsss=cmss10 at 7pt


\font\manual=manfnt \def\dbend{\lower3.5pt\hbox{\manual\char127}}


\def\boxit#1{\vbox{\hrule\hbox{\vrule\kern8pt
\vbox{\hbox{\kern8pt}\hbox{\vbox{#1}}\hbox{\kern8pt}}
\kern8pt\vrule}\hrule}}
\def\mathboxit#1{\vbox{\hrule\hbox{\vrule\kern8pt\vbox{\kern8pt
\hbox{$\displaystyle #1$}\kern8pt}\kern8pt\vrule}\hrule}}

\Title{ \vbox{\baselineskip12pt \hbox{hep-th/0009103}
\hbox{YCTP-P9-00 } \hbox{ITEP-2000/}
\hbox{} }} {\vbox{
\centerline{On Exact Tachyon Potential}
\bigskip
\centerline{in Open String Field Theory}
 \centerline{}}}
\medskip
\centerline{\bf A. A. Gerasimov $^{1}$, and S. L. Shatashvili
$^{2}$\footnote{*}{On leave of absence from St. Petersburg Branch
of Steklov Mathematical Institute,
 Fontanka,
St.
Petersburg,
Russia.}}

\vskip 0.5cm
\centerline{\it $^{1}$ Institute for Theoretical and Experimental
Physics,
Moscow,
117259, Russia}
\centerline{\it $^{2}$ Department of Physics, Yale University, New
Haven, CT  06520-8120 }
\vskip 1cm
In these notes
we revisit the tachyon lagrangian in the open string field theory
using background independent approach of Witten from 1992. We claim
that the tree level lagrangian (up to second order in derivatives
and modulo some class of field redefinitions) is given by $L =
e^{-T} (\partial T)^2 + (1+T)e^{-T}$.
 Upon obvious change of variables this leads to the potential energy $-
\phi^2 \log {\phi^2 \over e}$ with canonical kinetic term. This
lagrangian may be also obtained from the effective tachyon
lagrangian of the p-adic strings in the limit $p\rightarrow 1$.
Applications to the problem of tachyon condensation are discussed.

\medskip
\noindent

\Date{September 13, 2000}

\newsec{Introduction}

The problem of tachyon condensation attracted wide interest in the
recent string theory literature after the proposal of A. Sen
\ref\sen{A. Sen, Tachyon condensation on the brane antibrane
system,", hep-th/9805170, JHEP 9808 (1998) 012.}. In \ref\sentwo{A.
Sen, Universality of the Tachyon Potential, hep-th/9911116, JHEP
9912 (1999) 027.} it was argued that the tachyon potential takes
the form: \eqn\senz{V(T) = M f(T),} with $M$-mass of D-brane and
$f$
- universal function independent of the background where brane is
embedded. The conjecture of Sen states that $f(T)$ has a stationary
point (local minimum) at some $T=T_c < \infty$ such that
\eqn\con{f(T_c)
=
-1.} Thus total mass vanishes:
$M+V(T_c) = M (1+V(T_c))=0.$

Several arguments in favor of this conjecture were proposed.

It has been demonstrated in \sen, \ref\senzw{A. Sen and B.
Zwiebach, Tachyon condensation in string field theory,
hep-th/9912249, JHEP 0003 (2000) 002.} that sting field theory
action of Witten from 1986 \ref\witone{E. Witten, Noncommutative
Geometry And String Field Theory, Nucl. Phys. B268, 235 (1986)} and
 the systematic approximation scheme of Kostelecky and Samuel
\ref\kosts{V. A. Kostelecky and S. Samuel, The Static Tachyon
Potential in the Open Bosonic String Theory, Phys.Lett. B207 (1988)
169 .} may be successfully used for verifying the Sen's conjecture.
For example in {\it level zero} approximation:
\eqn\ks{f^0(t) = 2 \pi^2 (-{1 \over 2} t^2 + {1 \over 3} {t^3
\over r}), \quad \quad r = {4 \over 3\sqrt 3}.} This function has
local minimum at $t=t_c = r^3 = 0.456$ with the value $ f(t_c) = -
0.684$, which confirms the conjecture with 70\% of accuracy. This
approximation has been further improved and indeed it seems  that
the value of $f$ approaches $-1$ (though there is no obvious small
parameter expansion in level truncation method).

 In \ref\Gosh{D. Ghoshal and A. Sen, Tachyon Condensation and Brane
Descent Relations
 in p-adic String Theory, hep-th/0003278} the case of the p-adic strings
 was discussed.  It was
shown that the old results on the effective field theory
 of the tachyon   in the p-adic string theory \ref\freund{P. G. O Freund
and M. Olsen,
  Non-archimedian strings, Phys. Lett. B199 (1987) 186; P. G. O Freund and
E. Witten, Adelic
   string amplitudes, Phys. Lett B199 (1987) 191} could provide an
  interesting example for the discussion of the tachyon
  condensation.
  In \ref\olsenwitten{L. Brekke, P. G. O. Freund, M. Olsen and E. Witten,
  Non-archimedian string dynamics, Nucl. Phys. B302 (1988)365.} the
effective field theory
  of the tachyon which reproduce the
  tachyon scattering amplitudes in p-adic string theory was constructed.
  The lagrangian from \olsenwitten\ has the form:
  \eqn\plag{S(\phi)=
  \frac{1}{g^2}\frac{p^2}{p-1}\int d\sigma (-\frac{1}{2}\phi
p^{-\frac{1}{2} \triangle}\phi +
  \frac{1}{p+1}\phi^{p+1}),}
with the equations of motion: \eqn\pmotion{
p^{-\frac{1}{2}\triangle}\phi =
  \phi^{p}.}

In this paper we would like to approach the problem from the point
of view of two-dimensional theories living on the world-sheet of
the strings (for related recent discussions see  \ref\bcft{A.Sen,
SO(32) spinors of type I and other solitons on brane-antibrane
pair, JHEP 9809, 023 (1998), hep-th/9808141 ; V. A. Kosteletcky, M.
J. Perry and R. Potting, Off-shell structure of the string sigma
model, hep-th/9912243; J. A. Harvey, D. Kutasov and E. J. Martinec,
On the relevance of tachyon, hep-th/0003101, }) . For simplicity we
are dealing with the open strings in flat 26th dimensional space.

To verify the Sen scenario of tachyon condensation one needs to
have the definition of the string theory off-shell. It is well
known that the off-shell continuation of the theory suffers from
the field redefinition ambiguity (see for example \ref\tseyt{A. A.
Tseytlin, Phys. Lett. B264 (1991) 311}). We partially fix this
ambiguity by looking at the concrete recipe of the off-shell
definition of the string theory action proposed by Witten in
\ref\wbi{E. Witten, On Background Independent Open String Field
Theory, hep-th/9208027, Phys. Rev. D46 (1992) 5467-5473; E. Witten,
Some Computations in Background Independent Open-String Field
Theory, hep-th/9210065, Phys. Rev. D47 (1993) 3405-3410.} and
obtain the following effective lagrangian up to the second
derivatives in the tachyon field:
\eqn\eflagr{L =  e^{-T} (\partial T)^2 + (1+T)e^{-T}.}

It is natural to make a field redefinition to have  a derivative
term of standard form. This gives the following form of the tachyon
potential: \eqn\pot{V(\phi)=- \phi^2 \log {\phi^2 \over e}.}

Thus  we appear to obtain a lagrangian which shows up in many
interesting problems of theoretical physics for several decades
and string theory in particular.

Surprisingly it could be obtained from the effective lagrangian of
the tachyon in the p-adic string theory in the formal limit $p
\rightarrow 1$. Expanding the equation \pmotion\  around $p=1$ and
looking at the linear term we get:
 \eqn\old{\triangle \phi=2\phi \log\phi .}

Taking this lagrangian seriously leads to an interesting conclusion
of the fate of the tachyon in  open string theory. The potential
energy has the stable vacuum on the boundary of the configuration
space at $T=\infty$ or $\phi=0$. Around this  new vacuum the mass
of the tachyon excitations is infinite. This may be a manifestation
of the disappearance of the whole tower of open string fields at
this point and confirms some of the ideas from
\sen.

Obviously taking another off-shell continuation could change the
tachyon  lagrangian substantially. We hope however that the
 qualitative picture of the "new" tachyon vacuum will provide
 an interesting scenario for the tachyon condensation.

 Unexpected connection with the p-adic string also supports
 our result \eflagr.

\newsec{Background Independent Open String Field Theory}

 Let us first
remind the definition given by Witten for background independent
open string field theory \wbi.  This action is a functional of
boundary perturbations (corresponding couplings) of bulk CFT:

\eqn\wittwo{d S = < d\int_{\partial D} {\cal O} \{Q, \int_{\partial D}
{\cal O}\}>,}
where one considers the disk $D$ with boundary $\partial D$,
conformal field theory on $D$ (for simplicity we will take this CFT
to be just bosonic 26 dimensional string) perturbed with arbitrary
closed string operator $b_{-1}\int_C O$ integrated over the contour
$C$ with contour approaching the boundary of the disk $\partial D$;
$Q
= \int_C j_{BRST}$, and again contour $C$ approaches the boundary.
This also can be written in the BV formalism as \eqn\bv{dS = i_V
\omega,} where $\omega$ is odd-symplectic structure of BV formalism
and $V$ is a vector field that generates the symmetries of
$\omega$. This action has obvious property that it has critical
points exactly when boundary pertrubation is exactly marginal. The
general formula for this action has been found in \ref\shatash{S.
Shatashvili, Comment on the Background Independent Open String
Theory, hep-th/9303143, Phys. Lett. B311 (1993) 83-86; S.
Shatashvili, On the Problems with Background Independence in String
Theory, Preprint IASSNS-HEP-93/66, hep-th/9311177, Algebra and
Anal., v. 6 (1994) 215-226.} where it was demonstrated that:
\eqn\sh{S
=
- \beta^i\partial_i Z + Z,} with $Z$ -partition function and
$\beta^i$ - beta function for coupling $t^i$. For the simplest,
quadratic, boundary perturbation \eqn\quadr{ {\cal O} = c T, \quad
\quad T
= \frac{T_0}{2\pi} + \sum_i \frac{u_i}{8\pi} X_i^2,} the action was
computed in the second paper of \wbi:

\eqn\actone{ S = ( - \sum_j u_j {\partial \over \partial u_j} -
(T_0 + \sum_j u_j) {\partial \over \partial T_0} + 1) Z,} with
\eqn\part{ Z = e^{-T_0} \prod_i {\sqrt u_i}e^{\gamma u_i}
\Gamma(u_i).}

Simple calculations show that the asymptotic of the partition
function has the form: \eqn\asymptot{ Z=\sqrt{\frac{\pi}{4}}\int dX
e^{ - T_0 - \frac{u}{4} X^2}(1+O(u^2)).}

 In order to study the tachyon potential
it is enough to set $u_i=0$ and we find (up to the divergent
factor - in the limit $u \rightarrow 0$ there is no partition
function rather one has partition function per unit volume; for
discussion regarding this see \wbi):

\eqn\pot{V(T_0) = (1+T_0) e^{-T_0}.} This function gives exact
tree level tachyon potential when all other fields are set to zero
(corrections involve derivatives of tachyon and all other fields
which we have ignored at the moment) in a specific coordinate
system and regularisation dictated by world-sheet boundary sigma
model approach. It is obviously different than the potential
$Mf^0(t)$ from \ks, and although we can shift the potential by
$const = -1$ in order to have zero at $T=0$ and at the same time
get the value at the minimum $V(T_c)=-1$ this minimum is not
saturated and is at infinity $T=\infty$ which is very different
compared to Sen's conjecture.

Before we turn to the derivation of terms involving derivatives of
arbitrary tachyon field $T(X)$ in string field theory lagrangian we
would like to make some general remarks about the consequences of
the field redefinition for the tachyon lagrangian. We will discuss
only the effect of the field redefinition which contain at most the
two derivatives. Let us start with simple (non derivative) form of
the tachyon field redefinition:
\eqn\ann{ T
\rightarrow \widetilde{T}=f(T),} in the   action functional:
 \eqn\anseven{
S(T) = \int( h(T)(\partial T)^2 +V(T)+ \cdots ).} We have:
\eqn\aneigh{
 \widetilde{S}(T) =
 \int (h(f(T))(\frac{\partial f(T)}{\partial T} )^2(\partial T)^2 +
 V(f(T))+ \cdots
 ).}
 The equation for the  critical points of the potential is:
\eqn\annine{
 \frac{\partial f(T)}{\partial T}   \frac{\partial V}{\partial T}(f(T))=0
}
 It is clear that the possible "new" fixed points come from the zeros of
 the Jacobian of the field transformation (singular change of variables).
 Therefore  from \anseven\
 we conclude that in the "new" fixed points the coefficient
 (metric) in front of kinetic
 term is zero (if it was non-singular before
 the field redefinition).

However there is no reason to believe that we should not consider
more general tachyon field redefinition:\foot{Regarding recent
discussions of important physical role of field redefinition and
also full list of references on the subject of tachyon condensation
see \ref\senlast{A. Sen, Some Issues in Non-commutative Tachyon
Condensation, hep-th/0009038}.}

\eqn\anten{
 T \rightarrow \widetilde{T}=f(T)+g(T)(\partial  T)^2+\cdots ,}
 Under this transformation the new coefficient functions in the
 lagrangian are: \eqn\anel{
  \widetilde{V}(T)= V(f(T)),}
  \eqn\antw{\widetilde{h}(T)= h(f(T))(\partial f(T))^2+V'(f(T))g(T),}
  and we see that even if there was no kinetic term in original action
  it has been created: $L_{kin}=V'(f(T))g(T) (\partial T)^2$.
  For our potential in \pot\ this gives $L_{kin} = - g(T) T e^{-T}
(\partial T)^2$
  for $f(T)=T$.

In general nothing specific could be said about the new parameters
of the lagrangian. However adding some additional information from
boundary sigma model approach we could extract useful information,
because after fixing the regularisation scheme (treatment of
contact terms) in the definition via sigma model the action in
\wittwo\ is uniquely defined (up to a constant).

The kinetic term in the action defines the measure (metric in field
space) in second quantized path integral, so, for instance fixing
the boundary field theory analog of the Zamolodchikov metric to
have the simple exponential form $G_{TT}
\sim e^{-T}$ (up to same divergent factor which enters in
\pot\ and higher derivative corrections)
we come to the conclusion that kinetic term in our variables
$T$ is:
\eqn\kin{L_{kin} \sim e^{-T} (\partial T)^2.}
Indeed this turnes out to be the right answer and, as we show in a
moment, it follows directly from \wittwo, \sh; (it can be also
confirmed in standard sigma model approach \ref\gs{A. Gerasimov,
and S. Shatashvili, to be published.}).

Thus we would like to show that tachyon action up to two
derivatives in $T$ is:
\eqn\ansix{
 S(T)= \int e^{-T}((\partial T)^2+(T+1)).}
According \anten\ this action is a special case of the the whole
family of the tachyonic actions  \eqn\actionone{S_*(T)\sim \int
e^{-T}((1+g(T)T)(\partial T)^2+(T+1))} generated by the field
redefinitions \eqn\fred{T \rightarrow T - g(T) (\partial T)^2 +...}
with ... denoting the higher derivatives of tachyon field. In
addition one can also consider action of ordinary $Diff$ group.
Obviously we study only the perturbative class of equivalence of
this action (e.g. with non-perturbative $g(T) =
- {1 \over T}$ one can totally remove kinetic term from our action
\ansix). It seems that we should think about the action \ansix\ as
the one particular form which may be extracted from the
sigma-model.

Since we have the general expression for the action \sh\ we can
apply it to the case of boundary perturbation with only tachyon
field turned on: ${\cal O} = c(\theta) T(X(\theta))$. In this case
we know following expressions for $\beta$-function and partition
function:

\eqn\betaf{\beta^T(X) = (1 + 2 \Delta) T(X) + a_1(T)\partial T +
a_2(T) \partial^2 T + a_3(T) (\partial T)^2 +...,} \eqn\prt{Z =
\int dX e^{-T(X)} (1 + b(T) (\partial T)^2 + ...),}  with
\eqn\prop{a_1(0)=0, \quad a_2(0)=0, \quad a_3(0)=const, \quad
b(0)=const,} and all these coefficients are given by some concrete
power series expression in $T$ which can be determined from
boundary sigma model. In  fact by comparing \part\ with the
explicit formula \asymptot\ one could say more. Taking into account
the simple dependence of the action on the tachyon zero mode we
conclude that  $b$  does not depend on the tachyon field. Then
checking  the \prt\ against \asymptot\ gives us $b=0$.

Properties \prop\ just explain that \betaf, \prt\ are perturbative
expansions in tachyon and its derivatives with leading terms
determined by free field computations on world-sheet (note that for
the case studied in \wbi\ these properties are obviously
satisfied). Thus we have:

\eqn\ssh{S = - \int dX \beta^T(X) \partial_{T(X)} Z(T) + Z(T),}
with $\beta^T(X)$ and $Z(T)$ expressed in terms of \betaf\ and
\prt.

Interesting fact about the formula
\sh\ is that it encodes equations of motion in two ways: 1.
equations of motion are derived in stanard way from the action:
$dS=0$, 2. equations of motion derived in 1. are satisfied by zeros
of vector field in
\sh: $\beta^i = 0$.

After some simple algebra we find from
\sh\ and \betaf, \prt, \ssh:

\eqn\fff{S = \int e^{-T} [(2 + Q(T)) (\partial T)^2 + (T + 1)],}
where $Q(T)$ is linear combination of $a_i$'s and $b$ and their
derivatives with respect to $T$. \eqn\qu{Q(T) = - b + T(b - b') +
(a_2
- a_2') + a_3} Now we would like to show that $Q(0)=-1$.

 We use two ways of looking on equations of motion mentioned
above. Beta function equation in linear approximation is just free
tachyon equation $2\Delta T + T = 0$; the minimum of action \fff,
again in linear approximation, is given by $ 4\Delta T + 2Q(0)
\Delta T + T = 0$. These are consistent only if $Q(0) = -1$, so we
can write: $Q(T) = -1 + P(T) T$. We shall note that in quadratic
boundary perturbation \actone\ we had $a_1=a_2=0$ and $a_3= -1$.
This is consistent with the general expression \qu.

We conclude that string field theory action \wittwo, \sh, \fff\ is
in the equivalence class \fred\ of action \ansix, with $g(T) =
P(T)$. Explicit form of $P(T)$ depends on regularisation scheme
adopted in boundary sigma model. Simple  analysis of the dependence
on zero tachyon mode allows to fix the function $Q$ uniquely -
background independent open string theory naturally leads to the
constant function $Q(T)=-1$.

Interestingly, for the form \ansix\ on the equations of motion:
\eqn\mot{e^{-T}(2\Delta T - (\partial T)^2 +T) =0}
the whole action can be simply written as $\int e^{-T}$ (we just
integrate equation \mot\ over spacetime) and we learn that on-shell
action and partition function do indeed coincide (in the
approximation adopted throught this paper). Also we see that
together with beta function equation from
\betaf\ and partition function \prt\ we get: $b=0, a_1=a_2=0,
a_3= -1$. This completes our derivation of
\ansix.

This kinetic term for the tachyon has non standard form. In
particular it has zero at infinity in $T$ variables. Taking into
account the obvious field redefinition (element of remaining
$Diff$
):
 \eqn\ch{ e^{-T} = \phi^2}
 we obtain the final action announced in the Abstract:
 \eqn\fin{S \sim \int 4(\partial \phi)^2 - \phi^2 \log
{\phi^2 \over e}.} This field redefinition  is singular at
$T=\infty$ or same at $\phi=0$. We will comment about the relation
between the coordinates used in sigma model approach and
Chern-Simons string field theory at the end of the paper.

  We shall note that in the minimum
 of potential energy $V$ of derived action \ansix\ (which is at
$T=\infty$)
 metric $e^{-T}$ has zero in accordance
 with the conjecture of Sen. But difference is that in the variables
 $T$ of our approach  this minimum is at infinity
 and is never saturated (it is on the boundary of configuration space);
 it is not surprising that variable $T$ differs from $t$ used in \ks, but
certainly
 some relation can be established (see next section).

  The appealing property
 of this potential is that the effective mass of the tachyon
 excitations around the new minimum at $T=\infty$ is infinite.
 One could hope that similar mechanism gives the infinite masses
 to other open string excitations but this deserves further
 investigations.

\newsec{Some comments regarding the relation to level approximation
scheme in string field theory.}

Now  we would like to make some preliminary remarks on connection
of the field variables used in our sigma model analysis and the
field variables used in {\bf CS} string field theory for analysis
of the tachyon potential in the level approximation scheme \kosts
,\senzw. Obviously these  parameterizations of the string field
functional are quite
 different. In particular the gauge (BRST) transformation acts
 differently on these variables which is most obvious for U(1) gauge
 fields \foot{For recent discussion see e.g. \ref\david{
 J. R. David, U(1) gauge invariance from open string field theory,
 hep-th/0005085.}}.
 In the sigma model  approach the gauge transformation is independent on
background fields
 while in  {\bf CS} string field theory
it is linear in string fields.  The perturbative solution for this
 field redefinition enters in essential way in the
 level-approximation scheme of \kosts.

Below we outline  the construction of this field redefinition in
terms of the 2d functional integral. To define this field
redefinition we give the expression for the wave function
 in string field theory parameterized by the sigma model fields.

 Consider
 2d field theory on the disk with the bulk action $S_{Bulk}$ describing
the
 closed strings in the flat 26d background. Divide the boundary on two
equal parts
 $I_1$ and $I_2$.
 On $I_1$ we take the
     boundary conditions $X(\sigma)=X_*(\sigma)$ with some fixed
     $X_*(\sigma)$ playing the role of string wave
     function argument. On the other part $I_2$ of the boundary we
consider the  free
     boundary conditions but with the boundary action parameterized by the
sigma model variables:\eqn\BCFT{
     S_{bound}=-\int_{I_2} d \sigma (T(\sigma)+A_{\mu}\partial X^{\mu}
     +\cdots)}
     The proposed parameterizations of the open string wave function is
given by the following condition on variations: \eqn\psi{\delta
\Psi(X_*(\sigma))\sim \int
DX(\sigma)e^{S_{bound}+S_{bulk}}(\int_{I_2}d\sigma (\delta
T(X(\sigma)+\delta A_{\mu}(X(\sigma)) \partial
X^{\mu})(\sigma)+\cdots )}

One could formally integrate this
equation:\eqn\psii{\Psi(X_*(\sigma))\sim \int
DX(\sigma)e^{S_{bound}+S_{bulk}}+ \cdots} to get the wave function
up to integration constants. The main motivation for \psi\ comes
from the connection with the sigma model effective action. Note
that convolutions of these three wave functions
     with the Witten's  {\bf CS} open
     string product \witone\  gives the expression for the partition
     function on the disk with free boundary conditions and the
     boundary action \BCFT . This object is very close to the
     effective action for open string modes.
     Consider what gives this parameterization for the connection
     of the  constant tachyon modes $T_0$ in
     sigma model   with constant tachyon  mode $T^{CS}$ in {\bf CS} string
     field theory.
 We have:
      \eqn\condd{\frac{\partial T_0^{CS}}{\partial T_0}\sim
      e^{-\frac{1}{2}T_0}}
           (here $\frac{1}{2}$ comes from the integration of the tachyon
     over one-half of the boundary of the disk)
    Integrating this equation with the condition
    $T_0^{CS}(T_0=0)=0$ we get
            \eqn\conddd{T^{CS}_0 \sim e^{-\frac{1}{2}T_0}-1}
     In turn this gives the potential of the form:
     \eqn\condddd{V(T_0^{CS})\sim(1-2\log(T_0^{CS}+1))(T_0^{CS}+1)^2}
      Here the expansion around the "false"  vacuum with $T_0^{CS}=0$
      could be compared with  level approximation string field theory
results
      \kosts,\senzw. We plan to discuss the connection between
      sigma model and {\bf CS} open string theory variables more
      thoroughly in \gs.

\newsec{Final remarks}

We would like to conclude with few remarks not directly related to
the subject of these notes, but we believe they bring right flavor
to the discussion.

 The interesting example of the importance of
right choice of the fundamental fields is the theory of
non-critical strings, mainly $c=1$ model. The spectrum of the
theory consists of the one massless scalar field and a set  of
discrete states \ref\pol{D.J. Gross, I.R.Klebanov and M.J.Newman,
Nucl. Phys. B350 (1991),621; A. M. Polyakov,
 Self-tuning Fields and Resonant Correlations
  in 2d-Gravity, Mod. Phys. Lett. A6 (1991) 635.}.

   The
effect of these additional states on the S-matrix of the massless
particles reduces to the non-invertible linear field redefinition:
\eqn\polykov{\phi(p) \rightarrow
\frac{\Gamma(1-2p)}{\Gamma(2p)}\phi(p)}

All stringy phenomena (e.g. background independence) are hidden in
this field redefinition (for details and extensive list of
references see e.g. \ref\moor{P.Ginsparg, G.Moore, Lectures on 2D
Gravity and 2D String Theory, hep-th/9304011}

Finally note that according to the proposed scenario of the tachyon
condensation  the critical value of the tachyon field becomes
infinite.  We are going to discuss the implication of the
topological classification of the tachyon vacua (see \ref\witten{E.
Witten, D-branes and K-theory, JHEP 12, 019 (1998) hep-th/9810188})
elsewhere \gs.

{\bf Note added:} After this work was finished the paper
\ref\minzwi{J. Minahan and B. Zwiebach, Field Theory Models From
Tachyon and Gauge Field String Dynamics, hep-th/0008231.} appeared
were the potential in \fin\ has been studied as a model example
which mimics many expected properties of tachyon condensation. We
claimed that this potential is in fact related to exact one via
change of variables.

 {\bf Acknowledgements:} We would like to thank M. Douglas,
J. Maldacena, A.Morozov, N. Nekrasov, Y. Oz, So-Jong Rey and Erik
Verlinde for important discussions. S. Sh. like to thank CERN
theory division for hospitality during summer 2000 when this work
was finished. The research of A.G. is partially supported by RFBR
grant 98-01-00328 and the research of S. Sh. is supported by OJI
award from DOE.

\listrefs
\bye